\documentstyle[preprint,aps,eqsecnum]{revtex}

\input{psfig}
\begin{document}
\thispagestyle{empty}
\setcounter{page}{1}

\title{THE ELECTROMAGNETIC MASS DIFFERENCE OF PIONS 
AT LOW TEMPERATURE}

\author{Cristina Manuel}

\address{\baselineskip=12pt
Dpt. Estructura i Constituents de la Mat\`{e}ria\\
Facultat de F\'{\i}sica,
Universitat de Barcelona \\
Diagonal 647,
08028 Barcelona (SPAIN)}

\author{Nuria Rius}

\address{\baselineskip=12pt
Dpto. de F\'{\i}sica Te\'orica and IFIC \\
Centro Mixto Universidad de Valencia-CSIC \\
46100 Burjasot, Valencia (SPAIN)}

\maketitle

\thispagestyle{empty}
\setcounter{page}{0}

\begin{abstract}
$\!\!$We compute low temperature corrections to the electromagnetic mass 
difference of pions in the  chiral limit.  
The computation is done in a model independent way in the framework of 
chiral perturbation theory, using the background field method and
the hard thermal loop approximation. 
We also generalize at low temperature the 
sum rule  of Das et al.
We find that the mass difference between the charged and neutral pions 
decreases at low temperature $T$ with respect to the $T=0$ value. This is
so in spite of the fact that charged particles always get a thermal
correction to their masses 
of order $\sim eT$, where $e$ is the gauge coupling constant. Our result
can be understood as a  consequence
of the tendency towards chiral symmetry  restoration at finite
temperature.
\end{abstract}

\vfill

\hfill\break
\hbox to \hsize{ECM-UB-PF/98-14} 
\hbox to \hsize{FTUV/98-45} 
\hbox to \hsize{IFIC/98-46} 
\hbox to \hsize{June/1998}
\vskip-12pt
\eject

\setcounter{page}{1}

\baselineskip=15pt
\pagestyle{plain}

\section{INTRODUCTION}
\label{Intro}

At low energies the strong interactions are
successfully described in the framework of
chiral perturbation theory ($\chi$PT) \cite{We,GL}.
This theory only involves the low
energy modes of the QCD spectrum, such as the (pseudo) Goldstone
bosons of the spontaneously broken chiral symmetry. There are
eight (pseudo) Goldstone bosons, $(\pi's, K's, \eta)$, whose interactions
can be  understood in terms of symmetry considerations. A chiral
Lagrangian is expanded in derivatives of the Goldstone fields,  and
in the masses of the three light quarks, which break 
explicitly chiral symmetry.  Electromagnetic interactions
can also be included in a chiral Lagrangian, although they  break
the chiral symmetry explicitly.

The mass difference between the charged and neutral pions in the
chiral limit  (i.e. massless quarks)
was computed more than thirty years ago using a sum rule approach \cite{Das}.
With very general hypothesis, namely, soft pion theorems,  current algebra, 
Weinberg's sum rules \cite{Weinberg},
and vector meson dominance,  Das et al. 
reached to the theoretical prediction $(M_{\pi^\pm} - M_{\pi^0} )_{th} \sim 
5.0$ MeV. That value was in a remarkable good agreement with 
the experimental one $(M_{\pi^\pm} - M_{\pi^0} )_{exp} \sim 
4.6$ MeV. This computation allowed one to understand  this mass difference as
being due, essentially, to electromagnetic effects. This result was
also reproduced in the framework of $\chi$PT, by considering resonance
exchange within a photon loop \cite{deRafa}. Better theoretical estimates
of the electromagnetic mass difference of pions have since 
then been achieved \cite{Dono}.

In the  chiral limit Dashen's theorem \cite{Dashen}
states that the mass
difference between charged and neutral kaons should be the
same as the one of pions. Electromagnetic effects
can be parametrized in a chiral Lagrangian
in such a way that  Dashen's theorem is fulfilled in the  chiral limit 
\cite{deRafa}.
However, there is no good agreement between
the predicted electromagnetic mass difference and
the experimental value $(M_{K^\pm} - M_{K^0} )_{exp} \sim  -4.0$ MeV.
This is understood from the fact that the kaon's  mass difference receives
important QCD contributions of the order of the quark masses. On the other hand,
corrections to Dashen's theorem due to the effect of the quark masses
have been studied thoroughly in the literature 
(see, {\it e.g.} Refs. \cite{Urech} and
\cite{cor1} and references therein).

In this article we will compute the electromagnetic mass difference
of  pions and kaons  in the  chiral limit at low $T$.
 Exactly as it happens at $T=0$, we expect that our results
for those masses should be trusted for pions. To get a better estimate of 
$(M_{K^\pm}- M_{K^0}) (T)$ further corrections should be taken into account,
but we will not compute them here.
We will only consider low temperatures, that is, 
$T \ll  f_{\pi}$, where  $ f_{\pi}=92.4$ MeV is the pion decay constant.
This is the temperature regime where  the 
contribution of the lightest particles of the QCD spectrum
is the relevant one, since those of heavier states are
exponentially suppressed. Thus, the thermal contribution
of vector resonances will not be  taken into account in this article.

The computation will be first done in a model independent way in
the framework of $\chi$PT, using the
imaginary time formalism.
We will consider the lowest order chiral Lagrangian extended to include
electromagnetism, and obtain the one-loop thermal effective action for soft
modes with the help of the background field method. We will 
work in the hard thermal loop (HTL) approximation \cite{BP,HTL}.
This methodic approximation
allows us to  identify
the leading thermal corrections to any Feynman diagram, while ensuring
the gauge independence of the results.

Finally, we will generalize to low $T$  the sum rule computation of
Das et al. \cite{Das}.
Using the same kind of soft pion techniques as 
in the $T=0$ case, we will check our previous results for pions.
In this approach we will make use of the evaluation of 
thermal correlators of vector and
axial vector currents derived in Ref. \cite{Dey}.

We will make a big emphasis on the assumptions and
approximations involved in any of the two above mentioned approaches.
It  will then be clearly established that the standard soft pion techniques
at low $T$  in the  chiral limit
have their diagrammatic counterpart in the HTL approximation.

This paper is structured as follows. In Subsect. \ref{sec2A} we review
the $\chi$PT framework used in our computation. In Subsect. \ref{sec2B}
we obtain the one-loop thermal effective action for the soft Goldstone
bosons. The sum rule computation of Das et al \cite{Das} is reviewed
in Subsect.  \ref{sec3A}. This computation is then generalized at low
$T$ in Subsec. \ref{sec3B}. In Subsect. \ref{sec3C} we present the
numerical results, and we end with the conclusions in Sect. IV.
A general discussion on HTL's is presented in the Appendix.

\section{CHIRAL PERTURBATION THEORY COMPUTATION}
\label{sec2}

\subsection{Lowest Order Chiral Lagrangian with Electromagnetic Interactions}
\label{sec2A}

In this subsection we present a brief summary of $\chi$PT in the lowest order
extended to include electromagnetic interactions. We also explain how 
the background field method (BFM) is used in this specific case.
 We use natural units, so that
$\hbar = c = k_B =1 $.

At leading order the effective low energy Lagrangian
describing the physics of the (pseudo) Goldstone bosons
of the spontaneously broken chiral symmetry is, in Minkowski space-time
\cite{deRafa}

\begin{eqnarray}
{\cal L}_2 &  = & -  \frac{1}{4} F_{\mu \nu} F^{\mu \nu} - \frac {1}{2 a} (\partial_\mu A^\mu)^2
\label{eq:2.1}
\\
& + &   \frac{f_\pi^2}{4} \, {\rm Tr} \left(\nabla_\mu \Sigma^{\dagger} \nabla^\mu 
\Sigma\right) +  \frac{f_\pi^2}{4} {\rm Tr} \left( \chi^{\dagger} \Sigma + \chi  \Sigma^{\dagger} \right) 
 +  C \, {\rm Tr} \left( Q\Sigma Q \Sigma^{\dagger} \right) \nonumber \ ,
\end{eqnarray}
where $A_\mu$ is the photon field, $F_{\mu \nu}$ is the photon field strength, $a$ is
the gauge-fixing parameter, and $\Sigma$ is a $SU(N)$ unitary matrix written in terms
of the (pseudo) Goldstone bosons $\Phi$ as 
\begin{equation}
\Sigma = \exp { (i \Phi / f_\pi)} \ .
\label{eq:2.2}
\end{equation}
For $N=3$, $\Phi= \Phi^a \lambda^a $, where $\lambda^a $ are the Gell-Mann matrices, 
and in terms of the physical particles, one can write
\begin{equation}
\Phi = \sqrt{2}  \pmatrix{ \frac{\pi^0}{\sqrt{2}} + \frac{\eta_8}{\sqrt{6}} &  \pi^+ & K^+ \cr
\pi^- & - \frac{\pi^0}{\sqrt{2}} + \frac{\eta_8}{\sqrt{6}}  &  K^0 \cr
 K^- & {\bar K}^0 & - \frac{2\eta_8}{\sqrt{6}} \cr} \ .
\label{eq:2.3}
\end{equation}

The matrix $Q$ is the quark charge matrix, which
for the $N=3$ case reads
\begin{equation}
Q = \frac e3 \pmatrix{
2 &  &  \cr
 & -1 &   \cr
 &    & -1 \cr} \ ,
\label{eq:2.4}
\end{equation}
and $e$ is the electromagnetic coupling constant.

 The covariant derivative is defined as
\begin{equation}
\nabla_\mu \Sigma = \partial_\mu \Sigma - i (v_\mu + a_\mu + Q A_\mu) \Sigma + 
i  \Sigma (v_\mu - a_\mu + Q A_\mu) \ ,
\label{eq:2.5}
\end{equation}
$v_\mu$ and $a_\mu$ being external vector and axial vector hermitian and
traceless sources, respectively. One also defines 
\begin{equation}
\chi = 2 B (s+i p)  \ ,
\end{equation} 
where $B$ is a constant related to the quark condensate 
$\left< {\bar q} q \right> = -  f_\pi^2 \,  B \, [1 + O(m_q)]$, 
and $s$ and $p$ are scalar and pseudoscalar external sources,
respectively. 
The external scalar source can incorporate the mass matrix of quarks, 
which we put to zero since we will work in the chiral limit.  
In order to mantain the usual chiral counting, it is convenient to 
count the photon field as a quantity of order $O(1)$  and the 
electromagnetic coupling $e$ of $O(p)$, so the Lagrangian 
${\cal L}_2$ is of order $O(p^2)$ \cite{Urech}.

The parameter $C$ gives the purely electromagnetic part of the masses of the
charged pions and kaons in the chiral limit
\begin{equation}
M^2 _{\pi^{\pm}} = M^2 _{K^{\pm}} = \frac{2 e^2 C}{f_\pi ^2} + O (m_q)  \ .
\label{eq:2.6}
\end{equation}

The constant $C$ obeys a sum rule which relates its value to the
correlators of vector and axial vector currents (see Subsec. \ref{sec3A}).
The low energy constant $C$ is a model 
independent quantity, which enters naturally in the chiral Lagrangian.

To ensure the  $SU(N)_R \times SU(N)_L$ symmetry of the Lagrangian
(\ref{eq:2.1}) one introduces local spurions $Q^R (x)$ and $Q^L (x)$ instead
of the constant charge matrix,  Eq. (\ref{eq:2.4}). The Lagrangian (\ref{eq:2.1}) is
then modified so that the covariant derivative  becomes 
\begin{equation}
\nabla_\mu \Sigma = \partial_\mu \Sigma - i (F_\mu ^R + Q^R A_\mu) \Sigma + 
i  \Sigma (F_\mu ^L + Q^L A_\mu) \ ,
\label{eq:2.7}
\end{equation}
where, for convenience, we have defined
\begin{equation}
F_\mu ^R = v_\mu + a_\mu  \ , \qquad     F_\mu ^L = v_\mu - a_\mu  \ .
\label{eq:2.8}
\end{equation}
The last term in  Eq. (\ref{eq:2.1}) also becomes 
\begin{equation}
C \, {\rm Tr} \left( Q_R \Sigma Q_L \Sigma^{\dagger} \right) \ .
\label{eq:2.9}
\end{equation}

Under a $SU(N)_R \times SU(N)_L$ symmetry all fields transform as
\begin{mathletters}
\begin{eqnarray}
\Sigma' (x) & = & U_R (x) \Sigma (x) U_L  ^{\dagger} (x) \ , \\ 
Q'^l (x) & = & U_l (x) Q^l (x)  U_l  ^{\dagger}(x) \ , \qquad l = R, L \ , \\
\left(F_\mu ^R(x)  + Q^R (x) A_\mu (x)
\right)' & =& U_R (x) \left(F^R_\mu (x)  + Q^R(x)  A_\mu (x) \right )  U_R^\dagger (x) +
 i U_R (x) \partial_\mu
 U_R^\dagger (x) \\
\left(F_\mu ^L(x)  + Q^L(x)  A_\mu (x)\right )' & =& U_L (x)\left (F_\mu ^L(x)  + Q^L(x)  
A_\mu(x) \right)  U_L^\dagger (x) 
+ i U_L (x) \partial_\mu U_L^\dagger (x) \ , \\
\left(s(x) + i p(x)\right)' & = & U_R (x) \left(s(x) + i p(x)\right) U_L  ^{\dagger} (x) \ ,  \\
U_l (x)  &\epsilon & SU(N)_l \ , \qquad l = R, L \ . 
\label{eq:2.10}
\end{eqnarray}
\end{mathletters}

At this point one is ready to compute one-loop corrections to the lowest
order chiral Lagrangian with the background field method.
Let us recall that
at the end of the computation one has to put
  the spurions $Q_R (x) = Q_L (x) = Q$.

In the BFM all fields are split into classical and quantum pieces. 
The photon gauge field is split additively 
\begin{equation}
A_\mu (x) = {\bar A}_\mu (x) + \alpha_\mu (x) \ ,
\label{eq:2.11}
\end{equation}
where $ {\bar A}_\mu$ is the background field, and $\alpha_\mu$ is the
quantum one. The matrix containing the Goldstone fields is split 
multiplicatively
\begin{equation}
\Sigma (x) = \xi(x) \exp{(i \phi/ f_\pi)} \xi(x)  \ ,
\label{eq:2.12}
\label{split}
\end{equation}
where
\begin{equation}
\bar{\Sigma} (x) = \xi(x) \xi(x) \ ,
\label{eq:2.13}
\end{equation}
is the background or classical field, and $\phi$ is the quantum fluctuation.

In the spirit of the BFM, one expands the Lagrangian  keeping only
terms which are quadratic in the fluctuations
\begin{equation}
{\cal L}_2 = {\cal L}_2 ^{(0)} +  {\cal L}_2 ^{(2)} + \cdots 
\label{eq:2.14}
\end{equation}

We  will work in Euclidean space-time, so that we rotate the 
Minkowski Lagrangian ${\cal L}_2 $ to the Euclidean one
${\cal L}_{2,E}$. In Euclidean space-time one has \cite{Urech}
\begin{eqnarray}
\label{eq:2.15}
{\cal L}_{2,E} ^{(0)} &  = & -  \frac{1}{4}{\bar  F}_{\mu \nu} {\bar F}^{\mu \nu} - 
\frac {1}{2 a} (\partial_\mu {\bar  A}^\mu)^2
\\
& - &   f_\pi^2 \, {\rm Tr} ({\bar \Delta}_\mu)^2 -  \frac{f_\pi^2}{4} \,{\rm Tr} (M^+)
 - \frac{C}{4} \, {\rm Tr} \left( H^2_R - H^2_L \right) \nonumber \ ,
\end{eqnarray}
and
\begin{eqnarray}
\label{eq:2.16}
{\cal L}_{2,E} ^{(2)} & = &  \frac{1}{4}\, {\rm Tr}  ({\bar d}_\mu \phi)^2 - 
\frac{1}{4} {\rm Tr}\left([{\bar \Delta}_\mu, \phi] \right)^2 + \frac{1}{8}
{\rm Tr} (M^+ \phi^2)  
\\
&- &\frac{1}{4} \left( \partial_\mu \alpha_\nu - \partial_\nu \alpha_\mu \right)^2 - 
\frac{1}{2 a} (\partial_\mu \alpha^\mu)^2  +
\frac{f_\pi}{2} {\rm Tr} \left([{\bar \Delta}_\mu, \phi]\, H_R \right) \alpha^\mu
\nonumber 
\\
& - &\frac{C}{8 f^2_\pi} {\rm Tr} \left( [H_R + H_L, \phi] [H_R - H_L, \phi]\right) 
\nonumber 
\\
& -&\frac{f_\pi}{2} {\rm Tr} \left( {\bar d}_\mu \, \phi H_L\right) \alpha^\mu 
+ \frac {f_\pi ^2}{4} {\rm Tr} (H^2_L)
 \alpha^\mu \alpha_\mu 
\nonumber
\end{eqnarray}
where
\begin{mathletters}
\begin{eqnarray}
{\bar d}_\mu \phi& = &  \partial_\mu \phi +  [ {\bar \Gamma}_\mu, \phi] \ , \\
{\bar \Gamma}_\mu & = & \frac 12 \left( \xi^\dagger {\bar \nabla}_\mu ^R \xi +
 \xi {\bar \nabla}_\mu ^L 
\xi^\dagger \right) \ , \\
{\bar \Delta}_\mu & = &  \frac 12 \left( \xi^\dagger {\bar \nabla}_\mu ^R \xi - 
\xi {\bar \nabla}_\mu ^L 
\xi^\dagger \right) \ , \\
{\bar \nabla}_\mu ^l & = & \partial_\mu - i (F_\mu ^l + Q^l {\bar A}_\mu) \ , \qquad
 l = R, L  \ , \\
M^+ & = & \xi^\dagger \chi \xi^\dagger + \xi \chi^\dagger \xi \ , \\
H_R & = & \xi^\dagger Q_R \xi + \xi Q_L \xi^\dagger  \ , \\
H_L & = & \xi^\dagger Q_R \xi - \xi Q_L \xi^\dagger  \ ,  
\label{eq:2.17}
\end{eqnarray}
\end{mathletters}

We will work in the massless quark limit.
In this situation all particles at tree level are massless
except for the charged pions and kaons.

To obtain the one-loop effective action one has to integrate out the
quantum fluctuations $\phi$ and $\alpha_\mu$, while treating the background
fields as external sources. The functional integral can be done since
${\cal L}_{2,E}^{(2)}$ is only quadratic in the quantum fields. In a diagrammatic
approach, one should only consider diagrams with quantum fields running
inside the loop, while the background or classical fields only generate external
vertices. At $T=0$ the one-loop functional generated by ${\cal L}_{2,E}^{(2)}$
has been evaluated in Ref. \cite{Urech}, using dimensional regularization and renormalization
to deal with the ultraviolet  (UV) divergencies. In Ref. \cite{Urech} the 
next-to-leading order chiral Lagrangian ${\cal L}_4$ has also been constructed.
We will not need it at the order of the computation we are working.

\subsection{One-Loop Thermal Effective Action for Soft Modes}
\label{sec2B}

In this subsection we use the BFM to compute
one-loop thermal corrections to the low energy constant $C$ 
for soft modes. 
We will only consider thermal
corrections  and refer to \cite{Urech} for  a more general $T=0$
analysis.  Let us  mention that the UV divergencies which appear at 
$T=0$ and at finite $T$ are the same, so the renormalization procedure 
does not change at all in our analysis with respect to the $T=0$ one.

With respect to UV finite corrections in the  chiral limit,   and 
if we restrict our computation to soft modes, 
that is  momenta $P \ll T$, then one can ensure that the leading
thermal corrections, which are typically of order $T^2 /f^2_\pi$, are 
dominant with respect to the $T=0$ ones, which are of order $P^2 /f^2 _\pi$.
Therefore, if one neglects corrections of order  $P^2 /f^2 _\pi$,  
there are only  two expansion parameters in the  chiral limit.
One of them is the gauge coupling constant squared, $e^2$, and
the other is  the dimensionless quantity $ T^2 /f^2_\pi$.
We will assume that the two expansion parameters are equally important.

The computations  will be done using the imaginary time
formalism (ITF).  Feynman rules for propagators and vertices are straightforwardly 
derived from Eq. (\ref{eq:2.16}).
  We will denote Euclidean momentum with capital letters, so
$K^2 = k_0 ^2 + {\bf k}^2$. In ITF and for bosonic fields
$k_0 = 2\pi nT$
for integral $n$.  The standard notation for the thermal momentum measure
\begin{equation}
\int  \frac{d^4 K}{(2 \pi)^4} \equiv T \sum_{n=-\infty}^{n=\infty}
\int  \frac{d^3 k}{(2 \pi)^3} \ , 
\label{eq:2.18}
\end{equation}
will be  used throughout this article.

The leading thermal corrections to any
Feynman diagram when the external momenta is  {\it soft}, that is 
$\ll T$, arises when the momenta running inside the loop is {\it hard}, or
$\sim T$. Those diagrams are called {\it hard thermal loops} (HTL's) 
\cite{BP,HTL}.
There are systematic set of rules to extract from each Feynman
diagram the corresponding HTL. Those will be taken into account 
in the computation.

In this paper we are interested in finding the thermal corrections to the
masses of the soft charged pions and kaons. So we are looking for one-loop
corrections to the term in ${\cal L}_{2,E} ^{(0)}$
\begin{equation}
- \frac{C}{4} \, {\rm Tr} \left( H^2_R - H^2_L \right) \ .
\label{eq:2.19}
\end{equation}

Since we will only compute the leading thermal correction to the
effective action, and the masses of the charged quantum fields 
are  soft, we will neglect the masses of the quantum fields in the
propagators. This means that we are neglecting corrections of
order $M_{\pi^{\pm}}/T$ in the final answers. On the other hand,
considering those masses as soft quantities
means than we are assuming $T ^2 \gtrsim  C/f^2_\pi$.

There are only three Feynman diagrams to be evaluated, namely those
generated by the last three terms of Eq. (\ref{eq:2.16}), 
all of which give, in the HTL approximation, 
a tadpole type contribution at one-loop order.
We use dimensional regularization to deal with the UV divergencies.
In dimensional regularization and renormalization the $T=0$ contribution to 
the tadpole vanishes,
so we only  need to consider the thermal contributions
to each diagram.

The first diagram  we consider is a tadpole of quantum fields $\phi$.
The generators of $SU(N)$ $\lambda^a$ are normalized as ${\rm Tr}(\lambda^a
 \lambda^b)= 2 \delta^{a b}$, and $[\lambda^a, \lambda^b]= 2i f^{a bc} \lambda^c $, 
where $f^{a bc}$ are the structure constants of $SU(N)$. Then this tadpole
 gives the   correction to Eq. (\ref{eq:2.19})   
\begin{equation}
 \frac{C}{f_\pi ^2} \, f^{a b c} f^{a' b c} (H_R + H_L)^a (H_R - H_L)^{a'} 
 \int  \frac{d^4 K}{(2 \pi)^4} \frac{1}{K^2}  =
 C \, \frac{N}{24} \frac{T^2}{f^2_\pi} \, 
{\rm Tr} \left( H^2_R - H^2_L\right) \ .
\label{eq:2.20}
\end{equation}
In Eq. (\ref{eq:2.20}) we have used the $SU(N)$ relation $f^{a b c} f^{a' b c} = N \delta^{a
a'}$. 

Taking into account the relation
\begin{equation}
{\rm Tr} H^2_L = - \frac 12 \, {\rm Tr} \left(H^2_R - H^2_L\right) + {\rm Tr} 
\left(Q^2_R + Q^2_L\right) \ ,
\label{eq:2.21}
\end{equation}
one sees that there are two other diagrams which give corrections 
to  Eq. (\ref{eq:2.19}). One of them is a tadpole of the quantum gauge field, 
and the other is a diagram with a quantum boson $\phi$
and a quantum gauge field $\alpha_\mu$ circulating inside the loop.
We evaluate the two above diagrams in the Feynman gauge $a=1$.
However, since we  only retain the HTL of the diagram, one can
ensure that our results are gauge independent \cite{BP,HTL}.
We thus have
\begin{equation}
\frac{f^2_\pi}{4} \, {\rm Tr} H^2_L \left(
\int \frac{d^4 K}{(2 \pi)^4} \frac{\delta^{\mu \nu}\delta_{\mu \nu}}{K^2}
-  \int \frac{d^4 K}{(2 \pi)^4} 
\frac{( K + P)_\mu ( K + P)_\nu \delta^{\mu \nu}}
{K^2 (K+P)^2} 
 \right) \ .
\label{eq:2.22}
\end{equation}

Using the relation (\ref{eq:2.21})  we end up with 
\begin{equation}
- \frac{f^2_\pi}{8} \, {\rm Tr} \left( H^2_R - H^2_L\right) 
 \int \frac{d^4 K}{(2 \pi)^4} 
\frac{3}{K^2} = - \frac{f^2_\pi T^2} {32} \, {\rm Tr} \left( H^2_R - H^2_L\right) \ ,
\label{eq:2.23}
\end{equation}
plus an irrelevant constant term proportional to ${\rm Tr} 
\left(Q^2_R + Q^2_L\right)$.

The one-loop thermal effects for soft background fields involving the
${\bar \Gamma}_\mu$ and $ {\bar \Delta}_\mu$ fields arising from the two
first pieces of Eq. (\ref{eq:2.16}) have already been obtained in Refs. \cite{Man,PT}.
One can also easily compute the thermal correction to the term 
${\rm Tr} M^+$, just by evaluating a tadpole of $\phi$ fields, taking into
account the $SU(N)$ relation $\{\lambda^a, \lambda^b \} = \frac{4}{N} \delta^{ab} 
+2 d^{abc} \lambda^c$. 
Finally, one can write the complete one-loop thermal effective action
for soft modes as
\begin{eqnarray}
\label{eq:2.24}
{\cal Z}_2 = S_2 + \delta S_{2,T} & = &
\int d^4 x \, \left(-  \frac{1}{4}{\bar  F}_{\mu \nu} {\bar F}^{\mu \nu} - 
\frac {1}{2 a} (\partial_\mu {\bar  A}^\mu)^2 \right)
 \\
&- &\frac{N T^2}{12}\int \frac{d \Omega_{\bf q}}{4 \pi} \int d^4 x \,d^4 y\,
Tr \left({\bar \Gamma}_{\mu \lambda} (x) <x | \frac{Q^\mu Q_{\nu}}{- (Q \cdot
{\bar d})^2} |y> {\bar \Gamma}^{\nu \lambda} (y) \right)\nonumber  \\
\nonumber 
 &-&\int d^4 x \, \left(f_{\pi}^2 (T) \, Tr ({\bar \Delta}^2_\mu (x)) +\frac{C(T)}{4} \, {\rm Tr} \left( H^2_R(x) - H^2_L(x) \right) 
\right.
\\ 
 &-&  \left. 
\frac{f^2_\pi}{4} \left( 1- \frac{N^2-1}{12 N} \frac{T^2} {f_\pi ^2}
\right) {\rm Tr} (M^+ (x)) \right) \ ,
\nonumber
\end{eqnarray}
where  $Q^\mu$ is a light-like four vector (see Ref. \cite{Man}), 
and  \cite{GL2,BK,cool}
\begin{equation}
f_{\pi} (T) =  f_\pi \left( 1 - \frac{N}{24} \frac{T^2} {f_\pi ^2} \right) \ ,
\label{eq:2.27}
\end{equation}
and
\begin{equation}
{\bar \Gamma}_{\mu \nu} = \partial_\mu {\bar \Gamma}_\nu - 
 \partial_\nu {\bar \Gamma}_\mu
+  [{\bar \Gamma}_\mu, {\bar \Gamma}_\nu]  \ ,
\label{eq:2.25}
\end{equation}

\begin{equation}
C(T)  =  C \left( 1 - \frac{N}{6} \frac{T^2} {f_\pi ^2} \right) + \frac{T^2 f^2_\pi}{8} \ .
\label{eq:2.28}
\end{equation}

If we put the external sources $v_\mu = a_\mu = 0$, and $Q_R = Q_L =Q$,
the third term in Eq. (\ref{eq:2.24}) is just the HTL
 effective action \cite{efHTL}, which includes the Debye
mass for the electric field. 

With the above effective action one can read the thermal correction to the
quark condensate, just by performing a functional derivative of ${\cal Z}_2$ 
with respect to the external source $s(x)$.
It is given by  
\begin{equation}
\left< {\bar q} q \right>_T = \left< {\bar q} q \right>_{T=0} \left(1-\frac{N^2-1}{12 N} 
\frac{T^2} {f_\pi ^2}
\right) \ ,
\end{equation}
in  agreement with Ref. \cite{GL2}.

The value of the masses of the charged Goldstone bosons are obtained 
by finding the poles of the two-point functions of the axial current
(see Ref. \cite{GL} for details about the computation).
So one gets

\begin{equation}
M^2 _{\pi^{\pm}}(T)  = \frac{2 e^2 C(T)}{f_\pi ^2(T)}  \ .
\label{eq:2.29}
\end{equation}

In the low $T$ limit, and using the values of $C(T)$ and $f_\pi (T)$ above 
one then finds, at leading order,
\begin{equation}
M^2 _{\pi^{\pm}}(T)  = M^2 _{K^{\pm}} (T) \approx M^2 _{\pi^{\pm}} 
\left( 1 - \frac{N}{12} \frac{T^2} {f_\pi ^2} \right) + \frac{e^2 T^2}{4} \ .
\label{eq:2.29b}
\end{equation}

If one puts $C=0$, then the thermal masses of the charged bosons
agree with those computed in  Ref. \cite{Rebhan} for scalar QED. 
Temperature corrections to the electromagnetic mass difference 
of pions have also been considered in \cite{Kapusta}. We agree 
with the results of \cite{Kapusta}, except for the fact that 
thermal corrections of order $T^2/f_\pi^2$ were not considered 
there.

Finally, let us stress that in the massless quark limit Dashen's theorem
remains valid at finite $T$, as expected.

\section{SUM RULE APPROACH TO THE ELECTROMAGNETIC MASS 
DIFFERENCE OF PIONS}

\subsection{Computation at Zero Temperature}
\label{sec3A}

In this subsection we review the sum rule computation of the electromagnetic
mass difference of pions. We will follow closely the 
notation and conventions of Ref. \cite{Dono}, working in this subsection
in Minkowski space-time.  In the following subsection
we will generalize the computation at low $T$.

At $T=0$ one works with the assumption that $SU(2)_R \times SU(2)_L$ 
is a symmetry of the QCD Hamiltonian
which is spontaneously broken to   $SU(2)_{L+R}$.
The pions are then the associated Goldstone bosons. 

In the  chiral limit the neutral pion is massless, $M_{\pi^0} =0$,
while the charged pions get a mass from one-photon exchange 
contributions. The mass of the charged pions is given  by
\begin{equation}
M^2 _{\pi^{\pm}} = \frac{i e^2}{2} \int d^4 x \, \Delta^{\mu \nu} (x)
\left< \pi(p) \Big| {\cal T} \left ( J_\mu (x) J_\nu (0) \right) \Big| \pi(p) \right> \ ,
\label{eq:3.1}
\end{equation}
where $\Delta^{\mu \nu}$ is the photon propagator, and $J_\mu$ is the
electromagnetic current. 
In momentum space and in Feynman gauge, Eq. (\ref{eq:3.1}) becomes 
\begin{equation}
M^2 _{\pi^{\pm}} = \frac{i e^2}{2}  \int  \frac{d^4 q}{(2 \pi)^4} \, \frac{g^{\mu \nu}}
{q^2} \, T_{\mu \nu} (q^2, p \cdot q) \ ,
\label{eq:3.2}
\end{equation}
where 
\begin{equation}
T_{\mu \nu} (q^2, p \cdot q) = i \int d^4 x \, e^{i q \cdot x} \,
\left< \pi(p) \Big| {\cal T} \left ( J_\mu (x) J_\nu (0) \right) \Big| \pi(p) \right> \ .
\label{eq:3.3}
\end{equation}

Eq. (\ref{eq:3.3}) can be separated into non-contact (NC) and contact (C)
contributions \cite{Dono}. The last one refers to one which has both photons
interacting at the same vertex.

One can reduce the  non-contact contribution to the Compton scattering
amplitude to vacuum polarization functions \cite{Das,Dono}.
This is done using the soft pion theorem
\begin{equation}
\lim_{p_\mu \rightarrow 0} \left< \pi^a (p) \beta \Big| {\cal O} \Big| \alpha \right>
= - \frac{i}{f_\pi} \left< \beta \Big| [Q^a _5, {\cal O}] \Big| \alpha \right> \ ,
\label{eq:soft}
\end{equation}
where $\alpha$ and $\beta$ are arbitrary states, and $Q^a _5 $ is the axial
charge, and also the current algebra
\begin{equation}
[Q^a _5,  {\cal V}_\mu ^b] = i f^{a b c} {\cal A}_\mu ^c \ , \qquad
[Q^a _5,  {\cal A}_\mu ^b] = i f^{a b c} {\cal V}_\mu ^c  \ . 
\label{eq:algebra} 
\end{equation}
In Eq. (\ref {eq:algebra}) ${\cal V}_\mu ^a$ and ${\cal A}_\mu ^a$ are the vector current and 
axial vector current, respectively.
Thus, in the soft pion limit,
\begin{eqnarray}
\lim_{p_\mu \rightarrow 0} T_{\mu \nu}^{(NC)} (q^2, p \cdot q) & \equiv &
\frac{2 i} {f^2 _\pi} \left( \Pi_{\mu \nu} ^{V, 3} (q)- \Pi_{\mu \nu} ^{A,3}(q) \right)
\\
& = & \frac{2 i } {f^2 _\pi}
\int d^4 x \, e^{i  q \cdot x}
\left< 0 \Big| {\cal T}\left ({\cal V}_\mu ^3 (x) {\cal V}_\nu ^3 (0) - 
{\cal A}_\mu ^3 (x) {\cal A}_\nu ^3 (0) \right) \Big| 0 \right>  \ .
\nonumber
\end{eqnarray}

The above correlators are written in terms of spectral functions as
\begin{eqnarray}
\Pi_{\mu \nu} ^{V, 3} (q) & = &  i q^2 
\left( g_{\mu \nu} - \frac{q_\mu q_\nu}{q^2} \right)
\int_{0} ^{\infty} d s \, \frac {\rho_V (s)}
{q^2 - s} \ , 
\label{eq:3.4}\\
\Pi_{\mu \nu} ^{A,3}(q) & = &  i q^2
\left( g_{\mu \nu} - \frac{q_\mu q_\nu}{q^2} \right)
\int_{0} ^{\infty} d s \,  \frac {\rho_A (s)} {q^2 - s}
- i f_\pi ^2  \frac{q_\mu q_\nu}{q^2} \ .
\label{eq:3.5}
\end{eqnarray}

The contact term contribution to the Compton scattering amplitude is 
\cite{Dono}
\begin{equation}
T_{\mu \nu}^{(C)}  (q^2, p \cdot q) = 2 g_{\mu \nu} \ .
\end{equation}

Adding  the non-contact and contact contributions one gets
\begin{equation}
\lim_{p_\mu \rightarrow 0} T_{\mu \nu} (q^2, p \cdot q) 
= - \left( g_{\mu \nu} - \frac{q_\mu q_\nu}{q^2} \right)
\left( - 2 + \frac{2}{f^2 _\pi} \int_{0} ^{\infty} d s \left(
\rho_V (s)- \rho_A (s) \right) \frac{q^2}{q^2 - s} \right) \ .
\end{equation}

Therefore, in the mass formula one has
\begin{equation}
M^2 _{\pi^{\pm}} =  3 i e^2  \int  \frac{d^4 q}{(2 \pi)^4} \frac{1}{q^2} -
\frac{ 3 i e^2} {f^2_\pi} \int  \frac{d^4 q}{(2 \pi)^4}
\int_{0} ^{\infty} d s \, \frac {\rho_V (s) - \rho_A (s)}
{q^2 - s}  \ .
\label{eq:3.8}
\end{equation}
The above integrals are evaluated using dimensional regularization. 
Thus, the first term in Eq. (\ref{eq:3.8}) vanishes. 
The second one is logarithmically divergent, but the coefficient of 
the divergent piece vanishes due to the second Weinberg sum rule.
If one chooses a different UV regulator, there are also quadratic 
divergences which cancel due to the first Weinberg sum rule \cite{deRafa}.

The second term 
in Eq. (\ref{eq:3.8}) 
can be easily evaluated if
at this point one assumes that the spectral functions $\rho_V$ and 
$\rho_A$ are dominated by the vector and axial vector mesons \cite{Das}
\begin{eqnarray}
\rho_V (s) & = & F_{\rho}^2 \, \delta \left( s - M^2_\rho \right) \ ,
\label{eq:3.6}\\
\rho_A (s) & = & F_{A}^2 \, \delta \left( s - M^2_A \right) \ .
\label{eq:3.7}
\end{eqnarray}

Using the relations
\begin{equation}
F_{\rho}^2 \, M^2_\rho = F_{A}^2 \, M^2_A \ ,  \qquad
F_{A}^2 = F_{\rho}^2 - f^2_\pi  \ ,
\label{eq:3.9}
\end{equation}
which are derived from Weinberg's sum rules \cite{Weinberg}
\begin{eqnarray}
\int_{0} ^{\infty} d s \, \left( \rho_V (s) - \rho_A (s) \right) & = & f^2_\pi \ ,
\\
\int_{0} ^{\infty} d s \, s \left(\rho_V (s) - \rho_A (s) \right) & = & 0 \ ,
\end{eqnarray}
one finds
\begin{equation}
M^2 _{\pi^{\pm}} = - \frac{3 e^2 }{16 \pi^2}  M^2_\rho \, \frac{F_{\rho}^2 }{f^2_\pi} 
\, \ln{\frac {M^2_\rho}{M^2_A}} \ ,
\label{eq:3.10}
\end{equation}
which is the classical result of Das et al. \cite{Das}.

Better estimates of the correlator of vector and axial vector currents, not necessarily
assuming a narrow-width spectral function as in Eqs. (\ref{eq:3.6})-(\ref{eq:3.7}), 
have been considered in the literature \cite{Dono}, \cite{Peris}.

\subsection{Computation at Low Temperature}
\label{sec3B}

In this subsection we generalize the previous sum rule computation at low $T$,
working in Euclidean space-time and in the ITF.
At low $T$ one can assume that the main hypothesis that hold at $T=0$ are 
still valid, and that one only needs to consider small thermal corrections
to the previous computation.

At low $T$ chiral symmetry is still spontaneously broken, although at high 
enough $T$ it is supposed to be restored. We make the assumption
that at low $T$ one can use the same  kind of soft pion theorems than at $T=0$.
In the order of the computation we are working 
knowledge of the $T=0$ Weinberg's sum rules will be sufficient for us.
A generalization of Weinberg's sum rules at finite $T$ can 
be found in Ref. \cite{Shuryak}.

We start from Eq. (\ref{eq:3.1}), but now evaluated in a thermal bath at
equilibrium. We will follow the same steps as in Sect. \ref{sec3A} to 
reduce the thermal amplitude
\begin{equation}
\left< \pi (p) \Big| {\cal T} \left ( J_\mu (x) J_\nu (0) \right) \Big| \pi(p) \right>_T 
\label{eq:new3b1}
\end{equation}
to a thermal polarization function. Thus, we use the low $T$
generalization of the soft pion theorem  Eq. (\ref{eq:soft}) at leading 
order in the thermal correction, i.e.,
\begin{equation}
\lim_{p_\mu \rightarrow 0} \left< \pi^a (p) \beta \Big| {\cal O} \Big| \alpha \right>_T
= - \frac{i}{f_\pi(T)} \left< \beta \Big| [Q^a _5, {\cal O}] \Big| \alpha \right>_T \ ,
\label{eq:softlowT}
\end{equation}
and the current algebra. 
It is important to realize that Eq. (\ref{eq:softlowT}) only holds
at low $T$ and at leading order in $T^2/f^2_\pi$. At order $T^4/f^4_\pi$
there are two distinct pion decay constants \cite{cool}, due to the loss of 
Lorentz invariance, and therefore at that order Eq. (\ref{eq:softlowT}) 
would be modified.

The thermal amplitude Eq. (\ref{eq:new3b1}) is then written in terms of
the thermal correlators of vector and axial vector currents.
In Euclidean space-time, these correlators are given by 
\begin{eqnarray}
\Pi_{\mu \nu} ^{V, a} (K,T) & = &   \int d^4 x \, e^{i K \cdot x}
\sum_{n}
\left<n| {\cal T} {\cal V}_\mu ^a (x) {\cal V}_\nu ^a (0)
e^{(\Omega - H)/T} |n \right> \ , 
\label{eq:3.11} \\
\Pi_{\mu \nu} ^{A,a}(K,T) & = &  \int d^4 x \, e^{i K \cdot x}
\sum_{n} \left<n| {\cal T} {\cal A}_\mu ^a (x) {\cal A}_\nu ^a (0)
e^{(\Omega - H)/T} |n \right> \ , 
\label{eq:3.12}
\end{eqnarray}
where the sum is over the full set of eigenstates of the Hamiltonian $H$,
and $e^{-\Omega} = \sum_n \left<n | e^{-H/T} |n \right>$, and $K$ is 
the Euclidean momentum.
These thermal correlators have been computed in Ref. \cite{Dey}  
assuming that the lightest particles of the spectrum of $H$, that is,
the pions,
give the main contribution to the sum. Then, just by using the same
soft pion theorems that hold at $T=0$ and the current algebra,
and after integrating over the pionic thermal space, one gets for massless
pions
\begin{eqnarray}
\Pi_{\mu \nu} ^{V} (K,T) & = & 
\left( 1 - \frac{T^2}{6 f^2_\pi} \right) \Pi_{\mu \nu} ^{V} (K,T=0) +
\frac{T^2}{6 f^2_\pi} \,   \Pi_{\mu \nu} ^{A} (K,T=0) \ ,
\label{eq:3.13} \\
\Pi_{\mu \nu} ^{A}(K,T) & = & \left( 1 - \frac{T^2}{6 f^2_\pi} \right) \Pi_{\mu \nu} ^{A} (K,T=0) +
\frac{T^2}{6 f^2_\pi} \,   \Pi_{\mu \nu} ^{V} (K,T=0) \ .
\label{eq:3.14}
\end{eqnarray}
The two  above  low $T$ correlators can thus be expressed in terms of their 
$T=0$ values, although there is a mixing between the vector and axial vector 
ones. Let us stress that at finite $T$, and due to the loss of Lorentz 
invariance 
in the thermal bath, the general form of the thermal correlator of two 
currents should depend on more unknown functions
than the ones at $T=0$. However, 
at low $T$ one can assume that these correlators retain, approximately,
their covariant form \cite{Dey}.

Once we know the value of the above thermal correlators, 
we are ready to compute
the thermal corrections to the mass of the charged pions. 
We work in the ITF, where the thermal
momentum measure is the one written in Eq. (\ref{eq:2.18}).
From Eqs. (\ref{eq:3.13})-(\ref{eq:3.14}) we know that 
\begin{equation}
\Pi_{\mu \nu} ^{V} (K,T) - \Pi_{\mu \nu} ^{A}(K,T) = \left( 1 - \frac{T^2}{3 f^2_\pi} \right)
\left( \Pi_{\mu \nu} ^{V} (K,T=0) - \Pi_{\mu \nu} ^{A}(K,T=0) \right) \ .
\label{eq:3.15}
\end{equation}

In Feynman gauge, one then has
\begin{equation}
M^2 _{\pi^{\pm}} (T) = 3 e^2 
 \int  \frac{d^4 K}{(2 \pi)^4} \frac{1}{K^2} - \frac{3 e^2}{f^2_\pi (T)}
\left( 1 - \frac{T^2}{3 f^2_\pi} \right)  \int  \frac{d^4 K}{(2 \pi)^4}
\int_{0} ^{\infty} d s \, \frac{ \rho_V (s) - \rho_A (s)} { K^2 +s } \ .
\label{eq:3.16}
\end{equation}
The thermal contribution of the first term  in Eq. (\ref{eq:3.16})
does not vanish, as it happens for
the $T=0$ contribution in dimensional regularization. 
Notice that we have neglected the thermal factors in
front of the first term of Eq. (\ref{eq:3.16}) . Those terms would
yield thermal corrections of order $e^2 T^4/f^4_\pi$, which are
subleading at the order of the computation that we are
working. If one uses dimensional regularization to cure
the UV divergencies of the above integral, then the
quadratic divergencies of Eq. (\ref{eq:3.16})  vanish. If one uses
a different UV regulator then to check that the quadratic
divergencies which multiply the $T^2$ corrections cancel
one would need the thermal factors that we have neglected
in Eq. (\ref{eq:3.16}). To simplify our computation, we will work
with dimensional regularization, where Eq. (\ref{eq:3.16}) holds
exactly.

At this stage one can apply the hypothesis of vector meson dominance,
which is valid at $T=0$.  We should then compute the integrals
\begin{equation}
 \int  \frac{d^4 K}{(2 \pi)^4} \left( \frac{F_\rho ^2}{K^2 + M^2_\rho}
-\frac{F_A ^2}{K^2 + M^2_A} \right) \ .
\end{equation}
After performing the sum over Matsubara frequencies one then gets the
$T=0$ contribution (already evaluated in the previous subsection), plus
the pure thermal part.  However, since the masses of the vector 
resonances are much bigger than the temperatures we are considering,
$M_\rho, M_A \gg T$, the thermal corrections associated to those particles
are suppressed as $\sim \exp{(-M_{\rho, A} /T)}$.

Thus, collecting all the thermal corrections, we get
\begin{equation}
M^2 _{\pi^{\pm}} (T) = \left( 1 - \frac{T^2}{6 f^2_\pi} \right) 
M^2 _{\pi^{\pm}} + \frac{e^2 T^2}{4} \ ,
\label{eq:3.17}
\end{equation}
plus  subleading corrections.
Therefore, we have checked our formula (\ref{eq:2.29}) for the $N=2$  case.

It is worthwhile emphasizing that since the thermal corrections to the 
second term of Eq. (\ref{eq:3.16}) factorize, our results are also valid if
we use a better evaluation of the spectral functions than the ones 
given in Eqs. (\ref{eq:3.6})-(\ref{eq:3.7}). 

\subsection{Results}
\label{sec3C}

At zero temperature, the difference of the squared pion masses in 
the chiral limit is given by the result of Das et al., 
Eq. (\ref{eq:3.10}). Using the relations (\ref{eq:3.9}) to eliminate 
the parameters of the axial vector meson, $A$, it can be written as
\begin{equation}
M^2 _{\pi^{\pm}} - M^2 _{\pi^0}
= \frac{3 \alpha }{4 \pi}  M^2_\rho \, 
\frac{F_{\rho}^2 }{f^2_\pi} 
\, \ln{\frac {F^2_\rho}{F^2_\rho - f^2_\pi}} \ ,
\label{eq:3.20} 
\end{equation}
where $\alpha$ is the electromagnetic coupling constant.
Although this result was obtained in the chiral limit, it can be used 
to give an estimate of the electromagnetic
mass difference of pions. 
The difference of the squared pion masses is   
\begin{equation}
M^2 _{\pi^{\pm}} - M^2_{\pi^0} \sim 2 M_\pi \Delta M_\pi  \ .
\label{eq:3.21} 
\end{equation}
Taking the physical value of the pion mass $M_\pi = 135$ MeV, and  
the following values for the other parameters involved:
$F_\pi = 92.4 $ MeV, $M_\rho = 770$ MeV, $F_\rho=153$ MeV and 
$\alpha =1/137$, 
one gets $\Delta M_{\pi} = 4.8$ MeV, which  
is in very good agreement with the experimental value 
$(\Delta M_{\pi})_{exp} = 4.6$ MeV.

We thus expect that $M^2 _{\pi^{\pm}}(T) - M^2_{\pi^0}(T)$
at low temperature is also well approximated   
by the chiral limit calculation  Eq. (\ref{eq:3.17}), i.e., 
\begin{equation}
\Delta M_\pi^2 (T) \equiv M^2 _{\pi^{\pm}}(T) - M^2_{\pi^0}(T)
= \left( 1 - \frac{T^2}{6 f^2_\pi} \right) 
\Delta M_\pi^2 + \pi \alpha T^2 \ ,
\label{eq:3.22}
\end{equation}
where  $\Delta M_\pi^2 = M^2 _{\pi^{\pm}} - M^2_{\pi^0}$, given 
in Eq. (\ref{eq:3.20}).
The result is plotted in Fig. 1, as a function of the temperature.
Although our computation is only strictly valid for low T, we
have extrapolated it  up to $T \sim \sqrt{6} f_\pi
\sim 220$ MeV.
The dashed-dotted line in Fig. 1 represents the part proportional
to the electromagnetic mass difference at  $T=0$, $\Delta M_\pi^2$,
in Eq. (\ref{eq:3.22}), and we see that it decreases with $T$. The dashed
line corresponds to the typical thermal mass of charged particles,  which
grows with $T$, while the solid line is the full result.

\section{CONCLUSIONS}
\label{sec4}

The difference of correlators of vector and axial vector currents can be
taken as an order parameter of chiral symmetry breaking
\cite{EdR,Stern}. The sum rule
$$
f^2_\pi \, \delta^{ab} =  \frac i 3
\int d^4 x \, \left< 0 \Big| {\cal T} \left(
{\cal V}_\mu ^a (x) {\cal V} ^{\mu b} (0) - 
{\cal A}_\mu ^a (x) {\cal A} ^{\mu b} (0) \right) \Big| 0 \right>
$$
implies  the asymmetry of the vacuum, 
since the operator $ {\cal V}_\mu ^a (x) {\cal V} ^{\mu a} (0) - 
{\cal A}_\mu ^a (x) {\cal A}^{\mu a} (0)$ transforms as 
the irreducible representation $(3,3)$ of the symmetry group 
$SU(2)_L \times SU(2)_R$, and therefore if the vacuum were
invariant, the  above correlator 
should be zero. 
If chiral symmetry is restored at high $T$, this order parameter 
should then vanish.

Low $T$ computations of the vector and axial vector correlators show
a tendency towards chiral symmetry restoration \cite{Dey}. Although those
computations cannot be trusted at higher $T$,  they  give a good estimate 
of the critical temperature of the phase transition, $T_c \sim 160$ MeV,
which agrees with the values found in lattice computations \cite{lattice}.

The electromagnetic mass difference of pions depends on  the 
correlators of vector and axial vector currents \cite{Das}. In concordance
with the signs of chiral symmetry restoration at finite $T$,  one would 
naturally expect that this mass 
difference should decrease with $T$. This is what we have obtained for
low $T$, although it is not {\it a priori} obvious from our 
Eq. (\ref{eq:3.22}).
We have found a slow decrease of this mass difference with $T$
but only after a numerical analysis involving the $T=0$ values 
of the different parameters. It should be stressed that
even if chiral symmetry is restored at a certain
$T_c$, typical thermal effects occurring in the plasma of pions 
and photons will always generate a thermal mass for the charged 
particles going as
$\sim e T$.  Therefore, one can never expect that the electromagnetic
mass difference of pions should vanish at any high $T$.

Our computation was done in the limit of massless quarks.
In order to get a more trustable value of 
the mass difference $(M_{K^{\pm}}^2 - M_{K^0}^2)(T)$, one would need
to take into account also QCD corrections proportional to the quark masses,
apart from the pure electromagnetic correction computed here.
This was, however, beyond the scope of this article.


\acknowledgements

We are specially grateful to E. de Rafael for suggesting to us the interest
of carrying out this project and for very useful comments.
We also want to thank J.~I. Latorre, A. Pich and J. Prades
for useful discussions. 
This work was supported through founds
from the  CICYT projects  AEN95-0590 and AEN96-1718, from DGICYT under
grant PB95-1077 and from EEC under the TMR contract ERBFMRX-CT96-0090,
and from the CIRIT contact GRQ93-1047.

\vspace{10mm}
\appendix
\section{Hard Thermal Loops}
\label{appA}

In a gauge field theory at finite $T$ hard thermal loops are one-loop
diagrams which are as important as the tree amplitudes for soft momenta.
Soft denotes a energy scale $\ll T$. If $g$ is the gauge coupling constant, and
it is assumed that $g \ll 1$, then a soft scale is typically of the order 
$\sim g T$. For soft fields, then HTL's have to be resummed in order to
take  into account those one-loop effects consistently  \cite{BP}.

Braaten and Pisarski realized that HTL's only arise when all the external
momenta of a diagram are soft and the internal one is hard $\sim T$.
They were then able to establish some power counting rules to extract
from each Feynman diagram the corresponding HTL. 

In the non-linear sigma model 
 HTL's also appear for soft external momenta \cite{PT,Man}
In $\chi$PT in the  chiral limit without  gauge interactions 
the expansion parameter for thermal corrections 
is $T^2/f^2_\pi$ and it is this scale the one which allows
to define the scales soft $\sim \sqrt{T^2/f^2_\pi} \,T$  or hard $T$.
Resummation is not required in this case.

In the chiral Lagrangian with electromagnetic interactions included
there are then essentially
 two expansion parameters,  $e$ and $ \sqrt{T^2/f^2_\pi}$.
We have  assumed in this article that the two of them are of the same order.

In our case,
one can then apply the same power counting rules  derived in Ref. \cite{BP}.
Following Braaten and Pisarski's analysis for the Lagrangian (\ref{eq:2.16}),
 it is then easy to realize
that vertices which do not carry a momentum dependence do not
produce a HTL, except for tadpole-like diagrams. So apart from the tadpoles,
the only vertices which produce HTL's are
\begin{equation}
\frac 14 \, {\rm Tr } \left(\partial_\mu \phi \,[{\bar \Gamma}^\mu, \phi] \right) \ ,
\end{equation} 
and
\begin{equation}
- \frac{f_\pi}{2}\, {\rm Tr } \left(\partial_\mu \phi \, H_L \right) \alpha^\mu \ .
\end{equation}

\begin{figure} 
\begin{center}
\psfig{figure=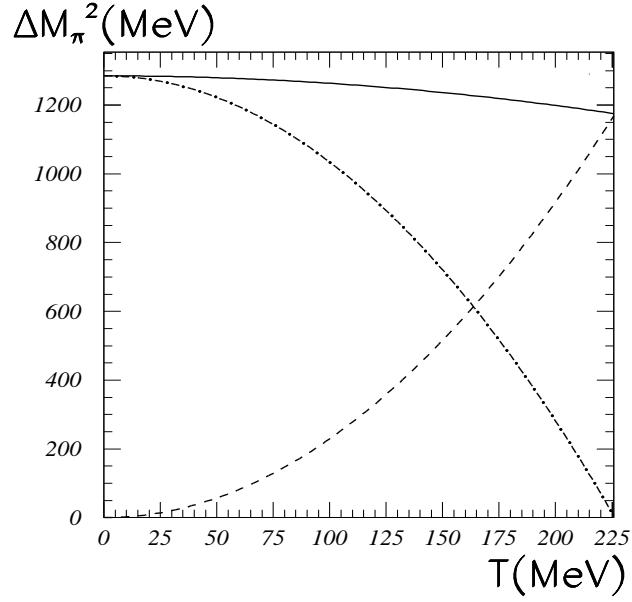,rwidth=2.5in,rheight=2.5in,width=5in,height=6in,bbllx=-30pt,bblly=-30pt,bburx=450pt,bbury=600pt}
\end{center} 
\vspace*{2cm}
\caption{Electromagnetic mass difference of pions 
$\Delta M_\pi^2 (T)$  
as a function of the temperature in the chiral limit. The dashed-dotted line
corresponds to the piece proportional to the electromagnetic mass difference
at $T=0$, $\Delta M_\pi^2$, the dashed line to the term proportional to 
$\alpha$ which grows with $T$, and the solid line is the full result}.
\end{figure}

\end{document}